\documentclass[conference,compsoc]{IEEEtran}
\IEEEoverridecommandlockouts

\usepackage{cite}
\usepackage{amsmath,amssymb,amsfonts}
\usepackage{algorithmic}
\usepackage{graphicx}
\usepackage{textcomp}
\usepackage{xcolor}
\usepackage{breakurl}
\def\BibTeX{{\rm B\kern-.05em{\sc i\kern-.025em b}\kern-.08em
    T\kern-.1667em\lower.7ex\hbox{E}\kern-.125emX}}

\usepackage{hyperref} 
\usepackage{makecell}
\usepackage{booktabs}
\usepackage{multirow}
\usepackage{float}
\begin{document}

\title{Fine-tuning Large Language Models for DGA and DNS Exfiltration Detection
}

\author{
    \IEEEauthorblockN{Md Abu Sayed\IEEEauthorrefmark{1}, Asif Rahman\IEEEauthorrefmark{1}, Christopher Kiekintveld\IEEEauthorrefmark{1}, Sebastian García\IEEEauthorrefmark{2}}
    
    \IEEEauthorblockA{\IEEEauthorrefmark{1}University of Texas at El Paso, El Paso, Texas 79968, USA}
    
    \IEEEauthorblockA{\IEEEauthorrefmark{2}CTU - Czech Technical University, Prague, Czech Republic}
    \IEEEauthorblockA{msayed@miners.utep.edu, arahman3@miners.utep.edu, cdkiekintveld@utep.edu, sebastian.garcia@agents.fel.cvut.cz}
}

\maketitle

\begin{abstract}
Domain Generation Algorithms (DGAs) are malicious techniques used by malware to dynamically generate seemingly random domain names for communication with Command \& Control (C\&C) servers. Due to the fast and simple generation of DGA domains, detection methods must be highly efficient and precise to be effective. Large Language Models (LLMs) have demonstrated their proficiency in real-time detection tasks, making them ideal candidates for detecting DGAs. Our work validates the effectiveness of fine-tuned LLMs for detecting DGAs and DNS exfiltration attacks. We developed LLM models and conducted comprehensive evaluation using a diverse dataset comprising 59 distinct real-world DGA malware families and normal domain data. Our LLM model significantly outperformed traditional natural language processing techniques, especially in detecting unknown DGAs. We also evaluated its performance on DNS exfiltration datasets, demonstrating its effectiveness in enhancing cybersecurity measures. To the best of our knowledge, this is the first work that empirically applies LLMs for DGA and DNS exfiltration detection.
\end{abstract}


\section{Introduction}

Malicious software or malware is a growing concern in cybersecurity\cite{sayed2022cyber,sayed2023honeypot}, with C\&C servers playing a central role in enabling attackers to control compromised systems, execute commands, and exfiltrate stolen data \cite{enisa}. One key technique botnets use to evade detection is DGAs, which create seemingly random domain names for establishing communication between bots and the C\&C server. These algorithms may be time-dependent or time-independent, generating multiple domains to obfuscate the server’s location. This technique, known as Domain Fluxing, complicates efforts to locate the C\&C server \cite{truong2016detecting}.

To combat botnets, one of the most effective countermeasures is identifying the C\&C domain through DNS request analysis. A technique called "sink-holing" reroutes bot requests to disrupt communication with the C\&C server. There are two primary approaches for detecting C\&C domains: signature-based (misuse) detection and anomaly detection. Signature-based detection relies on identifying known malicious domains, while anomaly detection focuses on recognizing domains that deviate from typical behavior. Both approaches can benefit from machine learning (ML), as classifiers can be trained on datasets containing benign and malicious domains to detect anomalies and misuse effectively \cite{al2019review}.

The data type and classification algorithms used greatly impact the effectiveness of DGA detection systems. Two major approaches are featureful and featureless methods. Featureful methods use predefined characteristics like the number of characters or n-gram distributions \cite{suryotrisongko2022robust}, while featureless methods treat domains as raw strings and leverage neural networks for classification \cite{zago2020scalable}. Recent literature tends to favor the featureless approach due to its simplicity and efficiency in detection, although it requires a large dataset for optimal performance \cite{cucchiarelli2021algorithmically}. 

LLMs are particularly well-suited for featureless DGA detection due to their ability to automatically extract relevant patterns and features from raw domain names, without requiring manual feature engineering. This is especially beneficial when dealing with large datasets, as LLMs can efficiently process and learn from extensive data thanks to tokenization and parallel processing techniques. By capturing the intricate patterns within domain name sequences, LLMs can effectively address the challenges associated with feature extraction, reducing the need for human intervention and improving detection accuracy in security tasks \cite{shestov2024finetuning}.

In this paper, we explore the application of LLMs for detecting DGA and DNS exfiltration attacks. Leveraging the generalizability of LLMs, we performed fine-tuning on a smaller dataset using Parameter-Efficient Fine-Tuning (PEFT) techniques. These methods allow us to adapt the LLM to specialized tasks while managing computational resources efficiently. We perform an extensive evaluation using a dataset containing 59 distinct real-world DGA malware families and normal domain data. The LLM significantly outperformed traditional natural language processing methods, particularly in identifying unknown DGAs. Additionally, its performance on DNS exfiltration datasets highlighted its potential to strengthen cybersecurity measures. To our knowledge, this is the first study to empirically apply LLMs for both DGA and DNS exfiltration detection, emphasizing the necessity of fine-tuning for these specific tasks.

We summarize our main contribution below:
\begin{itemize}
\item Validated the need for a fine-tuned LLM model and developed one for detecting DGA and DNS exfiltration attacks, followed by an in-depth evaluation of an extended dataset comprising domain names from 59 distinct real-world malware DGAs with varying generation schemes, as well as normal domains.
\item Evaluation of our fine-tuned LLM model shows that it not only surpasses established natural language processing techniques, such as N-Gram methods, on another dataset but also outperforms state-of-the-art models in detecting unknown DGAs.
\item Summary of key observations and final insights is
provided for the DNS exfiltration dataset.
\end{itemize}

\section{Related work}




The research on DGA detection has evolved from simple CNN models to more complex approaches incorporating multimodal information, lexical features, and advanced deep-learning techniques. The field has also progressed from binary classification to multi-class classification, enabling more granular detection of DGA types.

Early approaches by Catania et al. \cite{catania2019deep} conducted a thorough assessment and comparison of a Convolutional Neural Network (CNN) for detecting Domain Generation Algorithm (DGA) domains. The CNN, designed with minimal architectural complexity, was tested on a dataset comprising 51 distinct DGA malware families and normal domain traffic. Despite its simplicity, the model successfully identified over 97\% of DGA domains while maintaining a false positive rate of approximately 0.7\%.

Pei et al. \cite{pei2020two} introduce a novel approach for detecting DGA botnets by combining textual semantics and visual concepts, making it the first study to utilize multimodal information for this purpose. They propose a deep learning framework, TS-ASRCaps, which automatically learns multimodal representations from the data, eliminating the need for manual feature engineering. This innovative method paves the way for enhanced botnet detection using diverse data types.

Cucchiarelli et al. \cite{cucchiarelli2021algorithmically} developed a ML system that detects malicious domain names using lexical features, specifically 2-grams and 3-grams extracted from domain names. The system utilizes two primary metrics: the Jaccard index to evaluate the similarity between domain names and the Kullback–Leibler divergence to compare probability distributions between benign and malicious domains grouped by different DGAs. This method is applied to both binary and multiclass classification, making it the first study to integrate these metrics for domain name classification using both 2-gram and 3-gram features.


Namgung et al. \cite{namgung2021efficient} enhance DGA detection by extending deep learning-based methods from binary to multiclass classification, enabling the identification of specific DGA types. They propose a BiLSTM model and optimize it further with a CNN+BiLSTM ensemble, demonstrating superior performance on recent datasets. Chen et al. \cite{chen2023detection} focus on dictionary-based AGDs, using word segmentation and standard deviation features, and combining 3-gram and 1-gram sequence features for improved detection. They integrate these features into an end-to-end approach, achieving better accuracy in detecting both character-based and dictionary-based AGDs.

Enterprise networks are prime targets for cyber-attackers seeking to exfiltrate sensitive data via DNS channels. Jawad et al. \cite{ahmed2019real} develop and validate a real-time detection mechanism using a machine learning algorithm trained on benign DNS queries. Their approach tested on live 10 Gbps traffic streams from two organizations by injecting over a million malicious DNS queries, and they have made the tools and dataset publicly available.

\section{Methodology}





Large Language Models (LLMs) based on transformer architecture excel at content generation, but building one specifically for cybersecurity is resource-intensive. A more efficient approach is fine-tuning pre-trained LLMs with cybersecurity-specific datasets, leveraging their existing knowledge. This minimizes the need for extensive pre-training while enhancing capabilities like threat detection. To reduce computational demands, we use Parameter-Efficient Fine-Tuning (PEFT) techniques such as Low-Rank Adaptation (LoRA) \cite{hu2021lora} and Quantized LoRA (QLoRA) \cite{dettmers2024qlora}, which adjust only a small subset of model parameters while preserving most of the pre-trained ones.

LoRA \cite{hu2021lora} introduces a small, trainable submodule into the transformer architecture by freezing the pre-trained model weights and adding a low-rank decomposition matrix, significantly reducing the number of parameters needed for downstream tasks. Once training is complete, the low-rank matrix parameters are merged with the original model. QLoRA \cite{dettmers2024qlora} further optimizes this by using a 4-bit quantized version of the model, enabling efficient fine-tuning with minimal memory requirements while maintaining performance close to full fine-tuning. We use 4 pre-trained LLM model including BERT \cite{devlin2018bert}, Roberta \cite{liu2021robustly}, LLAMA3 \cite{touvron2023llama}, and Zephyr \cite{tunstall2023zephyr}.

In general, LLMs are trained on extensive amounts of unlabeled data using a unique methodology. This process involves taking tokens as input, predicting the next token in the sequence, and comparing it to the ground truth \cite{shestov2024finetuning}. This drives the next-token prediction loss. Specifically, for any given input sequence $\{x_1, x_2, \dots, x_N\}$ of length $N$, the model generates a probability distribution for the next token $P(x_{N+1}|x_1, x_2, \dots, x_N, \Theta) = p_{N+1} \in [0, 1]^v$, where $\Theta$ encompasses all model parameters and $v$ is the vocabulary size. By comparing this with the actual distribution—a one-hot encoded ground-truth token $y_{N+1} \in \{0, 1\}^v$—the cumulative cross-entropy loss can be minimized as:

\begin{equation}
    L_{ntp} = \sum_{n=1}^{N-1} y_{n+1} \log P(x_{n+1}|x_1, x_2, \dots, x_n, \Theta)
\end{equation}

However, when adapting an LLM for classification tasks, the objective needs to shift from next-token prediction to the classification goal. For an input sequence $\{x_1, x_2, \dots, x_N\}$, the standard pretraining objective would compute the loss across all predicted tokens, which may not be optimal for classification tasks like vulnerability detection, where the task is to classify the entire sequence rather than predicting the next token. To better align with classification tasks, we propose a loss function that focuses solely on the predicted probability of the input sequence $\{x_1, x_2, \dots, x_N\}$, which is matched against the true label $y$ using cross-entropy. The classification loss is defined as:

\begin{equation}
    L_{class} = - \log P(y|x_1, x_2, \dots, x_N, \Theta)
\end{equation}

In this context, $y$ denotes the correct class label, while $P(y|x_1, x_2, \dots, x_N, \Theta)$ represents the probability that the model assigns to the correct class. This approach ensures that weight updates during training are entirely driven by the classification task, with no interference from the generative pretraining objective. 








\section{Experimental Implementation}

In our experiments, we explored both binary and multi-class classification tasks. In the binary setting, we focused on differentiating
between benign domain names and those generated by
DGAs, regardless of the specific algorithm used for generation. We also consider different domain families for binary
classification. In the multi-class setting, we aimed to identify the exact DGA family responsible for generating the domain names. Identifying specific DGAs is crucial for gaining deeper insights into vulnerabilities and selecting the right countermeasures. Additionally, considering real-world scenarios
where new DGAs may emerge, it’s important for a classifier
to detect domains generated by unknown DGAs as they
surface. Our experiments address this challenge. Finally,
we evaluate our model performance on another dataset and
compare it with another state-of-the-art model \cite{cucchiarelli2021algorithmically}. We also share our findings on detecting DNS exfiltration attacks
and provide a detailed discussion on the dataset and the
application of LLMs in addressing this type of attack.
  
\subsection{Datasets}
\subsubsection{DGA Datasets}
Although there are various sources for benign and DGA-based domains, and several datasets used in research on detecting malicious domain names are available (such as AmritaDGA (Vinayakumar et al., 2019) and UMUDGA (Zago et al., 2020)), a standard benchmark dataset has not yet been established \cite{zago2020umudga}. In this paper, we utilize a combined dataset consisting of three different sources: one dataset for benign domains, which includes the Alexa Top 1 Million Sites collection of reputable domains \footnote{\url{https://www.kaggle.com/cheedcheed/top1m}}; and two datasets for DGA domains, sourced from Bambenek Consulting's malicious algorithmically-generated domains \footnote{\url{http://osint.bambenekconsulting.com/feeds/dga-feed.txt}} and the 360 Lab DGA Domains\footnote{\url{https://data.netlab.360.com/feeds/dga/dga.txt}}. Finally, the dataset is created by merging and shuffling these three datasets \footnote{\url{https://github.com/hmaccelerate/DGA_Detection/blob/master/data/mixed_domain.csv}}. In total, our DGA dataset comprises 1458863 domains that are DGA-based and 1000000 domains that are Alexa domains. The dataset comprises a total of 59 DGA domains and one benign domain. An overview of the used dataset is provided in Table ~\ref{tab:dga_data}.

We also use 25-DGA dataset \footnote{\url{https://github.com/chrmor/DGA_domains_dataset}} to compare our developed model performance with Cucchiarelli et al. \cite{cucchiarelli2021algorithmically}. The 25-DGA dataset comprises 25 distinct DGA families sourced from the Netlab Opendata Project repository \footnote{\url{data.netlab.360.com/dga}} and benign domain names from Alexa. The dataset includes a total of 675,000 domain names, evenly split between malicious and benign categories. 
Overall, this dataset has more than 50\% overlap with the DGA datasets.

\begin{table}[hbt!]
\caption{Overview of the DGA dataset.}
\label{tab:dga_data}
\centering
\scriptsize
\begin{tabular}{l r l}
\toprule
DGA & \# Domains & Examples \\

xshellghost & 1 & zsvubwnqlefqv.com \\
ccleaner & 1 & ab693f4c0bc7.com \\
madmax & 1 & www.nmwgcnadwz.net \\
blackhole & 2 & xlkaykasqozhuppr.ru \\

tofsee & 20 & dueduea.biz \\
tinynuke & 32 & \makecell[l]{08f1c1a243222466f50e192ade8e5e54\\.com} \\
omexo & 40 & \makecell[l]{023b2230f255816c166f4d665df0c704\\.net} \\
cryptowall & 94 & adolfforua.com \\
vidro & 100 & ahllpje.dyndns.org\\
proslikefan & 100 & adzvhm.ru\\
gspy & 100 & 01247dc1d13789b3.net \\
bamital & 104 & \makecell[l]{014d9e57888d4e2b783c438135d58a30.\\co.cc} \\
bedep & 178 & tfkgpjablr5q.com \\
hesperbot & 192 & nleflqnx.com \\

pykspa\_v2\_real & 200 & abprjmj.net \\
beebone & 210 & ns1.backdates0.biz \\
tempedreve & 255 & afcvuvgro.org \\
corebot  & 280 & 0k87re2wtynenwjy6k.ddns.net \\
fobber\_v1 & 298 & aaibkbnkncaxyjiph.net \\
fobber\_v2 & 299 & aajywwtpxk.com \\
conficker & 493 & abclnfnc.org \\
matsnu & 513 & ability-case.com \\
geodo & 576 & acigkycvyrdocbic.eu \\
fobber & 600 & aaibkbnkncaxyjiph.net \\
padcrypt & 744 & aaacofnekfkddcfn.ga \\
pykspa\_v2\_fake & 800 & adnkxyfrp.net \\

vawtrak & 812 & aberity.top \\
dircrypt & 821 & abnqumgstmnwpge.com \\
Volatile & 996 & adobeflashplatyerge.co.uk \\

chinad & 1,000 & 00e8k8h6aoq42bsc.org \\
cryptolocker & 1,000 & abrujanifnilt.org \\
pushdo & 1,680 & bacoqodaluc.kz\\
ramdo & 2,000 & aaaacqmeeoeumwey.org \\
qadars & 2,000 & 02kawmsa428q.org \\
P2P & 2,000 & aebynzcalfbbqjbpljvqsl.com\\

shifu & 2,554 & igmbesd.info\\
suppobox & 3,316 & toreking.ne \\
symmi & 4,320 & osutakfomaickee.ddns.net \\
locky & 5,163 & efvsxusdianhmrwnh.r \\
Cryptolocker & 6,000 & kxxtrmowmtth.net \\
nymaim & 6,309 & kjcplhuz.net \\
kraken & 6,958 & idxjoj.dynserv.com \\
dyre & 8,998 & \makecell[l]{tdc3e6d984803a757ff87b3ff158eb6c63\\.ws} \\
virut & 10,433 & bniifl.com \\
gameover & 12,000 & pfurgsvggyxkllfreivd.org\\
shiotob & 12,521 & 4ww5rdlc1b4bmz.net \\
pykspa & 14,215 & syasoaiq.net \\
ranbyus & 23,678 & nqniepiymsdjke.tw \\
simda & 31,044 & rypydal.info \\
murofet & 37,080 & wsoxolklejtslant.biz \\
qakbot & 40,000 & sortgymeuyeba.org \\
necurs & 40,960 & nsmljjaqlfbd.xxx \\
pykspa1 & 44,647 & uogoxwiugkeq.biz \\
ramnit & 57,728 & mdtyicvfelesdeh.com\\
Post & 66,000 & pqij0tpai87fswyqpw3u8bsh.net \\
tinba & 100,178 & vwwhinolkkme.in \\
rovnix & 179,980 & znukgz6o6fodhpv3vr.net \\
emotet & 286,816 & iqfindbnlvcfemde.eu \\
banjori & 439,423 & wyvzererwyatanb.com \\
\bottomrule
\end{tabular}
\end{table}

\subsubsection{DNS Exfiltration Dataset}
A substantial DNS dataset was captured from a live network environment, comprising over 50 million DNS queries. To protect privacy, IP addresses were anonymized. The data was carefully analyzed to extract features based on individual DNS requests and patterns across multiple requests. This processed dataset, reduced to approximately 35 million records, includes both normal DNS traffic and malicious exfiltration attempts. To increase the challenge of detection, a customized dataset with altered request patterns was generated \cite{vzivza2023dns}.

\subsection{Training Process}

The training process involves splitting the dataset, fine-tuning the LLM model using the training data, performing validation, and concluding with the testing phase.

\subsubsection{Dataset Split}

Our complete DGA dataset was divided into training, validation, and test sets with a split ratio of 30\%, 20\%, and 50\%, respectively. This stratified approach maintains the class distribution across all sets, ensuring a balanced representation. The split provides enough data for training, focuses on hyperparameter tuning, and reserves a large portion for testing to assess generalization. The training set (30\%) was used for fine-tuning the LLM, the validation set (20\%) for monitoring performance and preventing overfitting, and the test set (50\%) for unbiased final evaluation. For other tasks like binary, multi-class, and unknown domain classification, we used a smaller dataset (10k samples), split as 60\% training, 20\% validation, and 20\% testing, to test how well the LLM generalizes with a small amount of data.

\subsubsection{Fine-Tunning}

The fine-tuning process of LLM for text classification begins by loading a pre-trained model checkpoint configured for a binary classification task with two output labels. The tokenizer is modified to accommodate the input data by adding padding tokens where necessary. The dataset is then tokenized, transforming the text data into numerical inputs with truncation applied to limit the length of the sequences. To improve training efficiency, LoRA method is applied to BERT and Roberta, enabling updates to specific model components using fewer parameters. Additionally, we employ the QLoRA (4-bit quantization) technique to LLAMA and Zephyr as these models has 8 and 7 billion parameters respectively. Table ~\ref{tab:finetunned_model} represents the number of parameters in the pre-trained model we selected and the fine-tuned model we developed.


\begin{table}[hbt!]
\caption{Number of trainable parameters in different models}
\label{tab:finetunned_model}
\scriptsize
\centering
\begin{tabular}{l|r|r|r|r}
\hline
\makecell{Model} & \makecell{Base Model \\ Trainable \\ Parameters} & \makecell{Lora Based \\ Model\\ Trainable \\ Parameters} & \makecell{Qlora Based \\  Model\\ Trainable \\ Parameters} & \makecell{Fine-tunned \\ Model size} \\
\hline
BERT & 67,584,004 & \makecell{628,994\\ (0.93\%)} & -- & 3.5 MB \\
Roberta & 125,313,028 & \makecell{665,858\\ (0.53\%)} & -- & 5.8MB  \\
LLAMA3 & 7,518,572,544 & -- & \makecell{13,639,680\\ (0.18\%)} & 61 MB \\
Zephyr & 7,124,307,968 & -- & \makecell{13,639,680\\ (0.19\%)} & 55 MB \\
\hline

\end{tabular}
\label{tab:sample_table}
\end{table}


\subsubsection{Testing}

The testing process begins by evaluating the fine-tuned model on the test dataset using the trainer's prediction function, which generates predictions for the test samples. These predictions are then processed by determining the predicted class labels through the maximum probability. Various performance metrics are computed, including accuracy, precision, recall, and F1 score, which provide insight into the model's ability to correctly classify samples.
We evaluate the performance of the LLM model using test datasets of varying sizes, including 100k for large test data, 10k for binary classification across different DGA domains, 100k for multiclass classification, and 40k for unknown DGA domain classification.


\subsection{Evaluation Metrics}
We use accuracy, precision, recall, and F-measure metrics for evaluating the model's performance.

\textbf{Accuracy (Acc.):} The proportion of correctly classified samples out of the total samples, reflecting the confidence in the classification

\begin{equation}
    Acc. = \frac{TP + TN}{TP + FP + FN + TN}
\end{equation}

\textbf{Precision (Prec.):} The proportion of true positives (TPs) to the total of true positives (TPs) and false positives (FPs), indicating the confidence in the classification.

\begin{equation}
    Prec. = \frac{TP}{TP + FP}
\end{equation}

\textbf{Recall (Rec.):} The proportion of true positives (TPs) to the combined total of true positives (TPs) and false negatives (FNs), reflecting the completeness of the classification.

\begin{equation}
    Rec. = \frac{TP}{TP + FN}
\end{equation}

\textbf{F-Measure (F1):} The harmonic mean of precision and recall, providing an overall measure of the classification's performance.

\begin{equation}
    F1 = \frac{2*Prec. * Rec.}{Prec. + Rec.}
\end{equation}



\subsection{Experiments}
\subsubsection{DGA}

In our experiments, we investigate two distinct scenarios. The first scenario, widely adopted in the literature, involves both binary and multiclass classification of domains generated by known DGAs. In the binary classification, all DGA-generated domains are grouped into a single malicious class, and the goal is to distinguish them from benign domains using a large balanced test set of 100k samples. We further perform binary classification by grouping specific DGA-generated domains into balanced test sets of 10k samples each, focusing on 20 domains with over 5,000 examples each. In the multiclass classification experiment, each DGA family is assigned its class (19 classes total), along with an additional class for benign domains, resulting in 20 classes, each represented by 5k samples for performance testing.

Additionally, we tackle the identification of malicious domain names generated by unknown DGAs, a topic that is rarely addressed in the literature \cite{zago2020scalable,cucchiarelli2021algorithmically}. To simulate this scenario, we train the classifier in binary mode using malicious domains generated by various DGAs, excluding one domain class from the training data, and then test it on domains generated by the excluded DGA, using 20k benign and 20k excluded domains. We conduct 12 separate binary classification experiments, each considering a different DGA as the unknown class, focusing on 12 DGA domains that each have over 20k samples.

Binary, multiclass, and unknown DGA classifiers are fine-tuned using the full domain name, with 6k samples for training and 2k for validation. The training, validation, and test datasets are balanced, containing an equal distribution of benign and malicious domains.
The fine-tuned model is publicly accessible on Huggingface \footnote{\scriptsize{\url{https://huggingface.co/AbuSayed1/DGA-Meta-Llama-3-8B-FineTunnedModel}}} and our GitHub page \footnote{\scriptsize{\url{https://github.com/MDABUSAYED/StratosphereLinuxIPS/tree/LLM/modules/flowalerts/LLM}}}.

\subsubsection{DNS Exfiltration}

Before analyzing the DNS exfiltration data, we take specific steps to ensure that our model works with unbiased and balanced datasets. Our experiments are conducted on balanced DNS exfiltration data using two types of models: BERT and Hybrid BERT. The BERT model relies solely on textual data to detect DNS exfiltration attacks, while the Hybrid BERT model integrates both continuous and textual data for analysis.

\section{Results and Discussion}

In this section, we demonstrate the experimental results achieved under the various experimental settings explored in this paper. We present the performance of our fine-tuned model on our DGA dataset for both binary and multiclass classification of known DGAs. We also compare the effectiveness of our fine-tuned model against state-of-the-art models for classifying unknown DGAs. Finally, we evaluate the model's performance across the entire dataset and benchmark it against state-of-the-art models using different datasets, including the 25-DGA dataset.

We first evaluate the performance of the LLM model without any prior training on malicious or benign domains, a process known as testing the pre-trained model. The main motivation is that since LLM models are trained on vast corpora, they might perform well on this task; if the pre-trained model effectively distinguishes between benign and malicious domains, it could be used without further adjustments like fine-tuning for specific tasks. We use prompts such as '$[{'role':"user", "content":msg},]$', where msg is 'Is ' + 'www.msftncsi.com ' + 'a DGA domain? Please give me yes or no answer?'. However, the accuracy of the pre-trained LLM for binary classification is below 50\%, indicating the need to work with a fine-tuned LLM model. 

\begin{table*}[hbt!]
\caption{Results of the binary classification task using our method on the DGA dataset (tested with large data, 100k). We report the overall accuracy and, for each class (benign and malicious), precision, recall, and F-1 score. Best results are highlighted in bold.}
\label{tab:known_bc_bm}
\centering
\scriptsize
\begin{tabular}{c c c c | c c c | c c c | c c c}
\toprule
\multicolumn{4}{c}{BERT-FT} & \multicolumn{3}{c}{Roberta-FT} & \multicolumn{3}{c}{LLAMA3-FT} & \multicolumn{3}{c}{Zephyr-FT} \\
\midrule
Class & Prec. & Rec. & F1 & Prec. & Rec. & F1 & Prec. & Rec. & F1 & Prec. & Rec. & F1 \\
\midrule
Benign & 1.0 & 0.977 & 0.988 & 1.0 & 0.983 & 0.991 & 1.0 & \textbf{0.991} & \textbf{0.996} & 1.0 & 0.985 & 0.993  \\
Malicious & 1.0 & 0.967 & 0.983 & 1.0 & 0.977 & 0.988 & 1.0 & \textbf{0.981} & 0.99 & 1.0 & \textbf{0.981} & 0.990  \\
\midrule
Accuracy & & & 0.972 & & & 0.98 & & & \textbf{0.986} & & & 0.983 \\
\bottomrule
\end{tabular}
\end{table*}

\begin{table*}[hbt!]
\caption{Results of the binary classification task using our method on the DGA dataset across different DGA families. We present the overall accuracy, precision, recall, and F-1 score. Best outcomes are highlighted in bold.}
\label{tab:known_bc_df}
\centering
\scriptsize
\begin{tabular}{c c c c | c c c | c c c | c c c}
\toprule
\multicolumn{4}{c|}{BERT-FT} & \multicolumn{3}{c|}{Roberta-FT} & \multicolumn{3}{c|}{LLAMA3-FT} & \multicolumn{3}{c}{Zephyr-FT} \\
\midrule
Class & Prec. & Rec. & F1 & Prec. & Rec. & F1 & Prec. & Rec. & F1 & Prec. & Rec. & F1 \\
\midrule
Cryptolocker & 0.980 & 0.996 & 0.988 & 0.984 & 0.991 & 0.988 & \textbf{0.991} & \textbf{0.999} & \textbf{0.995} & 0.983 & 0.998 & 0.991 \\
Nymaim & 0.978 & 0.873 & 0.922 & 0.982 & 0.856 & 0.915 & \textbf{0.990} & 0.930 & \textbf{0.961} & 0.982 & \textbf{0.940} & 0.960 \\
Kraken & 0.979 & 0.935 & 0.957 & 0.983 & 0.920 & 0.950 & \textbf{0.991} & \textbf{0.976} & \textbf{0.983} & 0.982 & 0.958 & 0.970 \\
Dyre & 0.977 & 0.839 & 0.903 & 0.984 & \textbf{1.000} & \textbf{0.992} & \textbf{0.990} & 0.958 & 0.974 & 0.982 & 0.952 & 0.967 \\
Virut & 0.952 & 0.394 & 0.558 & 0.966 & 0.450 & 0.614 & \textbf{0.983} & 0.518 & 0.678 & 0.970 & \textbf{0.560} & \textbf{0.710} \\
Gameover & 0.980 & \textbf{1.000} & 0.990 & \textbf{0.984} & \textbf{1.000} & 0.992 & 0.991 & \textbf{0.999} & \textbf{0.995} & 0.983 & \textbf{1.000} & 0.991 \\
Shiotob & 0.980 & 0.978 & 0.979 & 0.984 & 0.988 & 0.986 & \textbf{0.990} & \textbf{0.999} & \textbf{0.995} & 0.983 & 0.998 & 0.990 \\
Pykspa & 0.978 & 0.909 & 0.943 & 0.983 & 0.908 & 0.944 & \textbf{0.990} & 0.959 & \textbf{0.975} & 0.983 & \textbf{0.961} & 0.972 \\
Ranbyus & 0.980 & 0.997 & 0.989 & 0.984 & 0.998 & 0.990 & \textbf{0.991} & \textbf{1.000} & \textbf{0.995} & 0.983 & 0.999 & 0.991 \\
Simda & 0.969 & 0.616 & 0.753 & 0.979 & 0.764 & 0.859 & \textbf{0.989} & 0.822 & 0.898 & 0.981 & \textbf{0.882} & \textbf{0.929} \\
Murofet & 0.980 & 0.999 & 0.990 & 0.984 & 0.999 & 0.991 & \textbf{0.991} & \textbf{1.000} & \textbf{0.995} & 0.983 & \textbf{1.000} & 0.991 \\
Qakbot & 0.980 & 0.987 & 0.984 & 0.984 & 0.985 & 0.985 & \textbf{0.991} & \textbf{0.997} & \textbf{0.994} & 0.983 & \textbf{0.997} & 0.990 \\
Necurs & 0.979 & \textbf{0.945} & 0.962 & 0.983 & 0.953 & 0.968 & \textbf{0.990} & \textbf{0.982} & \textbf{0.986} & 0.982 & 0.937 & 0.959 \\
Pykspa1 & 0.980 & \textbf{0.924} & 0.951 & 0.983 & 0.927 & 0.954 & \textbf{0.990} & 0.963 & \textbf{0.976} & 0.983 & \textbf{0.967} & 0.975 \\
Ramnit & 0.980 & 0.973 & 0.977 & 0.984 & 0.967 & 0.975 & \textbf{0.990} & \textbf{0.991} & \textbf{0.991} & 0.983 & \textbf{0.991} & 0.987 \\
Post & 0.980 & \textbf{0.999} & 0.989 & 0.984 & \textbf{0.999} & 0.993 & \textbf{0.991} & \textbf{0.999} & \textbf{0.995} & 0.983 & \textbf{0.999} & 0.991 \\
Tinba & 0.980 & 0.994 & 0.987 & 0.984 & 0.992 & 0.988 & \textbf{0.990} & \textbf{0.998} & \textbf{0.995} & 0.983 & 0.996 & 0.990 \\
Rovnix & 0.980 & 0.999 & 0.989 & 0.984 & 0.999 & 0.992 & \textbf{0.992} & \textbf{1.000} & \textbf{0.995} & 0.983 & \textbf{1.000} & 0.991 \\
Emotet & 0.980 & \textbf{1.000} & 0.990 & 0.984 & 0.999 & 0.992 & \textbf{0.990} & \textbf{1.000} & \textbf{0.995} & 0.983 & \textbf{1.000} & 0.991 \\
Banjori & 0.980 & 0.980 & 0.980 & 0.984 & \textbf{0.993} & \textbf{0.988} & \textbf{0.991} & 0.984 & 0.987 & 0.983 & 0.985 & 0.984 \\
\midrule
Accuracy & & & 0.95 & & & 0.96 & & & \textbf{0.973} & & & 0.970 \\
\bottomrule
\end{tabular}
\end{table*}

\subsection{Known DGA: Binary Classification}

Table ~\ref{tab:known_bc_bm} displays the performance of our developed model large balanced test datasets. Overall, LLAMA3-FT outperforms other LLM models, achieving an accuracy of 98.6\%.  
Table ~\ref{tab:known_bc_df} illustrates our model’s performance on binary
classification across various domain types. LLAMA3-FT remains the top-performing model, followed by Zephyr-FT, Roberta-FT, and BERT-FT. However,
the accuracy here is slightly lower than in Table ~\ref{tab:known_bc_bm}. This
decrease is because we averaged the accuracy
across different domain types, with 20 DGA domain types
having equal data amounts, which was not the case previously.


\subsection{Known DGA: Multi-class Classification}

\begin{table*}[hbt!]
\caption{Results of the multi-class classification task on the DGA dataset. Best results are marked in bold.}
\label{tab:known_mc_df}
\centering
\scriptsize
\begin{tabular}{c c c c | c c c | c c c | c c c}
\toprule
\multicolumn{4}{c|}{BERT-FT} & \multicolumn{3}{c|}{Roberta-FT} & \multicolumn{3}{c|}{LLAMA3-FT} & \multicolumn{3}{c}{Zephyr-FT} \\
\midrule
Class & Prec. & Rec. & F1 & Prec. & Rec. & F1 & Prec. & Rec. & F1 & Prec. & Rec. & F1 \\
\midrule
Nymaim & \textbf{0.530} & 0.441 & 0.481 & 0.476 & 0.504 & 0.490 & 0.452 & 0.655 & \textbf{0.535} & 0.446 & \textbf{0.715} & 0.550 \\
Kraken & 0.944 & 0.843 & 0.891 & 0.945 & 0.835 & 0.887 & 0.932 & 0.842 & 0.884 & \textbf{0.955} & \textbf{0.845} & \textbf{0.896} \\
Dyre & 0.998 & \textbf{1.000} & \textbf{0.999} & \textbf{0.999} & \textbf{1.000} & \textbf{0.999} & \textbf{0.999} & \textbf{1.000} & \textbf{0.999} & \textbf{0.999} & \textbf{1.000} & \textbf{0.999} \\
Virut & 0.745 & \textbf{0.981} & 0.847 & 0.752 & 0.970 & 0.848 & \textbf{0.903} & 0.964 & 0.933 & 0.882 & 0.918 & \textbf{0.940} \\
Gameover & 0.451 & 0.176 & 0.253 & 0.446 & 0.445 & 0.185 & \textbf{0.488} & 0.708 & 0.579 & 0.481 & \textbf{0.906} & \textbf{0.628} \\
Shiotob & 0.953 & 0.902 & 0.927 & 0.966 & 0.885 & 0.923 & 0.987 & \textbf{0.917} & \textbf{0.951} & \textbf{0.997} & 0.905 & 0.949 \\
Pykspa & 0.338 & \textbf{0.278} & \textbf{0.305} & 0.286 & 0.113 & 0.162 & 0.380 & 0.196 & 0.258 & \textbf{0.399} & 0.196 & 0.263 \\
Ranbyus & 0.718 & 0.780 & 0.747 & 0.847 & 0.662 & 0.743 & 0.852 & \textbf{0.799} & \textbf{0.824} & \textbf{0.894} & 0.764 & 0.823 \\
Simda & 0.871 & 0.912 & 0.891 & 0.846 & 0.919 & 0.881 & 0.860 & \textbf{0.949} & 0.902 & \textbf{0.898} & 0.945 & \textbf{0.921} \\
Murofet & 0.660 & \textbf{0.752} & 0.703 & 0.704 & 0.718 & 0.711 & 0.804 & 0.611 & 0.694 & \textbf{0.840} & 0.675 & \textbf{0.749} \\
Qakbot & 0.585 & 0.483 & 0.529 & \textbf{0.623} & 0.478 & 0.541 & 0.588 & 0.577 & 0.582 & 0.565 & \textbf{0.615} & \textbf{0.589} \\
Necurs & 0.915 & 0.728 & 0.811 & \textbf{0.963} & 0.701 & 0.812 & 0.911 & \textbf{0.783} & \textbf{0.842} & 0.929 & 0.770 & \textbf{0.842} \\
Pykspa1 & 0.542 & 0.637 & 0.586 & 0.523 & \textbf{0.854} & \textbf{0.649} & 0.550 & 0.605 & 0.576 & \textbf{0.576} & 0.630 & 0.601 \\
Ramnit & 0.589 & 0.535 & 0.560 & \textbf{0.594} & 0.626 & 0.610 & 0.549 & \textbf{0.755} & \textbf{0.636} & 0.572 & 0.695 & 0.627 \\
Post & 0.520 & 0.842 & 0.643 & 0.521 & \textbf{0.914} & \textbf{0.663} & \textbf{0.541} & 0.278 & 0.368 & 0.538 & 0.038 & 0.071 \\
Tinba & 0.640 & 0.791 & 0.707 & 0.569 & 0.810 & 0.669 & \textbf{0.820} & 0.922 & \textbf{0.868} & 0.751 & \textbf{0.924} & 0.828 \\
Rovnix & 0.966 & 0.956 & 0.961 & 0.967 & 0.971 & 0.969 & 0.992 & \textbf{0.987} & \textbf{0.989} & \textbf{0.999} & 0.978 & 0.988 \\
Emotet & 0.935 & 0.999 & 0.966 & 0.936 & \textbf{1.000} & 0.967 & \textbf{0.979} & 0.994 & \textbf{0.986} & 0.975 & 0.992 & 0.983 \\
Banjori & 0.946 & 0.983 & 0.964 & 0.945 & 0.991 & 0.967 & \textbf{0.987} & \textbf{0.994} & 0.971 & 0.978 & 0.987 & \textbf{0.982} \\
Benign & 0.932 & 0.806 & 0.865 & 0.946 & \textbf{0.783} & 0.857 & 0.955 & \textbf{0.907} & \textbf{0.930} & \textbf{0.956} & 0.901 & 0.928 \\
\midrule
Accuracy & & & 0.741 & & & 0.742 & & & \textbf{0.77} & & & 0.769 \\
\bottomrule
\end{tabular}
\end{table*}


Table ~\ref{tab:known_mc_df} summarizes the results of the multi-class classification on our DGA datasets, with the best performances highlighted in bold. LLAMA3-FT and Zephyr-FT showed nearly identical results, followed by Roberta-FT and BERT-FT. The overall accuracy was 77\%, significantly lower than the 95\% reported for the 25-DGA and UMDGA datasets \cite{cucchiarelli2021algorithmically}. This lower accuracy is likely due to the complexity of multiclass classification with over 20 classes and the limited data used (6k data for fine-tuning multiclass problem).

\subsection{Unknown DGA: Binary Classification}

\begin{table*}[hbt!]
\caption{unknown DGA domain results. For each unknown DGA (first column), we report the results obtained by the four classifiers.}
\label{tab:unknown_bc_df}
\centering
\tiny
\begin{tabular}{c c c c c | c c c c | c c c c | c c c c | c}
\toprule
\multicolumn{5}{c|}{BERT-FT} & \multicolumn{4}{c|}{Roberta-FT} & \multicolumn{4}{c|}{LLAMA3-FT} & \multicolumn{4}{c|}{Zephyr-FT} & N-Gram \\
\midrule
Class & Prec. & Rec. & F1 & Acc. & Prec. & Rec. & F1 & Acc. & Prec. & Rec. & F1 & Acc. & Prec. & Rec. & F1 & Acc. & Acc.\\
\midrule
Simda & 0.957 & 0.526 & 0.679 & 0.751 & 0.911 & 0.210 & 0.340 & 0.60 & \textbf{0.981} & \textbf{0.622} & 0.761 & \textbf{0.805} & \textbf{0.984} & 0.541 & 0.70 & 0.766 & \textbf{0.986} \\
Ranbyus & 0.969 & 0.995 & 0.982 & 0.982 & 0.969 & 0.998 & 0.983 & 0.983 & \textbf{0.984} & 0.999 & \textbf{0.992} & \textbf{0.992} & \textbf{0.980} & 0.999 & 0.989 & 0.990 & \textbf{0.996} \\
Murofet & 0.970 & 0.998 & 0.984 & 0.984 & 0.979 & 0.998 & 0.989 & 0.989 & \textbf{0.990} & 0.999 & \textbf{0.994} & \textbf{0.995} & 0.981 & \textbf{1.0} & 0.990 & 0.990 & 0.990 \\
Qakbot & 0.965 & 0.991 & 0.978 & 0.978 & 0.971 & 0.986 & 0.978 & 0.978 & \textbf{0.988} & \textbf{0.998} & \textbf{0.993} & \textbf{0.993} & \textbf{0.990} & 0.995 & \textbf{0.992} & \textbf{0.992} & - \\
Necurs & 0.965 & 0.959 & 0.961 & 0.962 & 0.965 & 0.934 & 0.949 & 0.950 & \textbf{0.987} & \textbf{0.985} & \textbf{0.986} & \textbf{0.990} & 0.981 & 0.940 & 0.960 & 0.961 & 0.990 \\
Pykspa1 & 0.961 & 0.882 & 0.920 & 0.923 & 0.975 & 0.869 & 0.919 & 0.923 & 0.980 & \textbf{0.968} & \textbf{0.974} & \textbf{0.974} & \textbf{0.988} & 0.940 & 0.964 & 0.965 & \textbf{0.983}\\
Ramnit & 0.964 & 0.974 & 0.969 & 0.969 & 0.981 & 0.946 & 0.963 & 0.964 & 0.984 & \textbf{0.995} & \textbf{0.990} & \textbf{0.990} & \textbf{0.988} & 0.988 & \textbf{0.988} & 0.988 & \textbf{0.994} \\
Post & 0.965 & \textbf{1.0} & 0.982 & 0.982 & 0.972 & \textbf{1.0} & \textbf{0.986} & \textbf{0.986} & \textbf{0.991} & 0.978 & 0.985 & 0.985 & 0.973 & 0.994 & \textbf{0.986} & \textbf{0.986} & - \\
Tinba & 0.964 & 0.994 & 0.979 & 0.979 & \textbf{0.988} & 0.984 & 0.986 & 0.986 & \textbf{0.990} & 0.999 & \textbf{0.994} & \textbf{0.994} & 0.960 & \textbf{0.999} & 0.982 & 0.981 & 0.993 \\
Rovnix & 0.971 & 0.991 & 0.985 & 0.985 & 0.976 & 0.999 & 0.988 & 0.988 & \textbf{0.991} & \textbf{1.0} & \textbf{0.996} & \textbf{0.996} & 0.983 & 0.999 & 0.991 & 0.991 & 0.987\\
Emotet & 0.981 & 0.992 & 0.990 & 0.991 & 0.975 & 0.999 & 0.987 & 0.987 & \textbf{0.992} & \textbf{1.0} & \textbf{0.996} & \textbf{0.996} & 0.970 & \textbf{1.0} & 0.984 & 0.985 & 0.995 \\
Banjori & 0.967 & 0.730 & 0.832 & 0.852 & 0.968 & 0.716 & 0.823 & 0.846 & 0.984 & \textbf{0.960} & \textbf{0.970} & \textbf{0.970} & \textbf{0.993} & 0.740 & 0.848 & 0.867 & -   \\

\bottomrule
\end{tabular}
\end{table*}

The classification results for the unknown DGA setting are presented in Table ~\ref{tab:unknown_bc_df}. Each row of the table begins with the DGA used to test the classifiers, representing the unknown DGA. Generally, classifying unknown DGAs is more challenging than known DGAs. We evaluated 12 unknown DGA domains, 9 of which align with the unknown DGA experiments in \cite{cucchiarelli2021algorithmically}. Our LLAMA3-FT model outperforms in certain unknown DGA types, achieving higher accuracy than the model developed by Cucchiarell et al. \cite{cucchiarelli2021algorithmically}, for murofet, tinba, rovnix, and emotet domain. In every instance, our LLAMA3-FT model outperformed the other classifiers, achieving notably strong results, such as for banjori (97\%), pykspa (97.4\%), and simda (80.5\%). Overall, these findings indicate that our developed model is effective in recognizing DGA variants over time.

We achieved an overall average accuracy of 97.2\% across 12 unknown DGA domains, which is slightly lower than the accuracy of the binary classifier for known DGA domains. This difference is due to evaluating performance on previously unseen DGA families and comparing it with known DGA binary classifiers. Moreover, this comparison highlights the generalizability of our developed LLM model.

\subsection{Performance on Full Dataset and Compare with Other Datasets}

\begin{table*}[hbt!]
\caption{Performance on Full Dataset and Compare with Datasets}
\label{tab:dataset_comparison}
\centering
\tiny
\begin{tabular}{c c c c c | c c c c | c c c c | c c c c | c}
\toprule
\multicolumn{18}{c}{\textbf{a) Our DGA Dataset}} \\
\midrule
\multicolumn{5}{c|}{BERT-FT} & \multicolumn{4}{c|}{Roberta-FT} & \multicolumn{4}{c|}{LLAMA3-FT} & \multicolumn{4}{c|}{Zephyr-FT} & N-Gram \\
\midrule

Train & \multicolumn{4}{c|}{2556 samples per second} & \multicolumn{4}{c|}{1484 samples per second} & \multicolumn{4}{c|}{10 samples per second} & \multicolumn{4}{c|}{8 samples per second} & -  \\

\makecell{Validation} & \multicolumn{4}{c|}{6367 samples per second} & \multicolumn{4}{c|}{3865 samples per second} & \multicolumn{4}{c|}{33 samples per second} & \multicolumn{4}{c|}{26 samples per second} & -  \\

\makecell{Inference} & \multicolumn{4}{c|}{619 samples per second} & \multicolumn{4}{c|}{388 samples per second} & \multicolumn{4}{c|}{63 samples per second} & \multicolumn{4}{c|}{59 samples per second} & -  \\

Criteria & Prec. & Rec. & F1 & Acc. & Prec. & Rec. & F1 & Acc. & Prec. & Rec. & F1 & Acc. & Prec. & Rec. & F1 & Acc. & Acc.\\
\midrule
Test & 0.994 & 0.994 & 0.994 & 0.992 & 0.992 & 0.992 & 0.992 & 0.991 & \textbf{0.997} & 0.996 & \textbf{0.996} & \textbf{0.996} & 0.996 & \textbf{0.997} & \textbf{0.996} & 0.995 & - \\

\multicolumn{18}{c}{\textbf{b) 25-DGA Dataset, full domain}} \\
\midrule
\multicolumn{5}{c|}{BERT-FT} & \multicolumn{4}{c|}{Roberta-FT} & \multicolumn{4}{c|}{LLAMA3-FT} & \multicolumn{4}{c|}{Zephyr-FT} & N-Gram \\
\midrule
Criteria & Prec. & Rec. & F1 & Acc. & Prec. & Rec. & F1 & Acc. & Prec. & Rec. & F1 & Acc. & Prec. & Rec. & F1 & Acc. & Acc.\\
\midrule
Test & 0.971  & 0.974 & 0.973 & 0.973  & 0.959 & 0.958 & 0.959 & 0.959 & 0.981  & 0.990 & 0.986 & 0.986  & 0.983 & 0.985 & 0.984 & 0.984 & 0.995 \\

\multicolumn{18}{c}{\textbf{b) 25-DGA Dataset, no TLD}} \\
\midrule
\multicolumn{5}{c|}{BERT-FT} & \multicolumn{4}{c|}{Roberta-FT} & \multicolumn{4}{c|}{LLAMA3-FT} & \multicolumn{4}{c|}{Zephyr-FT} & N-Gram \\
\midrule
Criteria & Prec. & Rec. & F1 & Acc. & Prec. & Rec. & F1 & Acc. & Prec. & Rec. & F1 & Acc. & Prec. & Rec. & F1 & Acc. & Acc.\\
\midrule
Test &  0.948 & 0.937 & 0.942 & 0.942 & 0.939 & 0.935 & 0.937 & 0.937 & \textbf{0.974} & \textbf{0.976} & \textbf{0.975} & \textbf{0.975} & 0.967 & 0.973 & 0.971 & 0.971 & 0.992 \\

\bottomrule
\end{tabular}
\end{table*}

Table ~\ref{tab:dataset_comparison} presents fine-tuned model performance on the full dataset and compares our developed model performance with another model over different datasets. Our LLAMA3-FT outperforms other model in terms
of performance over different metrics such as precision, recall,
F-1, and accuracy. However, the LLAMA3-FT and Zephyr-FT model number of sample processing over train, validation, and inference is much less compared to BERT-FT and Roberta-FT as they have large number of parameters. The performance of our LLAMA3-FT model surpasses that of the model developed by Cucchiarell
et al. \cite{cucchiarelli2021algorithmically} on the UMDGA datasets. Additionally, our developed fine-tuned model outperforms previous work \cite{pei2020two,chen2023detection,catania2019deep}. It is worth noting that this comparison is not on the same dataset but they have overlap.

We compared the performance of our fine-tuned model with natural language processing techniques, such as N-Gram methods \cite{cucchiarelli2021algorithmically}, using the 25-DGA dataset over both the full domain name and without the top-level domain. The dataset was divided into 30\% for training, 20\% for validation, and 50\% for testing. Our fine-tuned model's accuracy was slightly lower, by about 0.9\%, compared to the N-Gram model. This difference is mainly due to their use of 10-fold cross-validation, allowing their model to utilize the entire dataset.

\subsection{LLM Model Performance on DNS Exfiltration Attack}
The DNS exfiltration dataset includes both regular requests and exfiltrations carried out using DNSExfiltrator and Iodine tools \cite{vzivza2023dns}. However, there are two types of malicious DNS requests: the first involves real benign exfiltrations performed by AV products, specifically ESET and McAfee. We take this into account to ensure a balanced dataset. Some requests from the attack tools are repeated, so we need to address this to avoid working with duplicated data and ensure our datasets remain unbiased. Additionally, since the attacks are generated in a simulated environment and all share the domain '.dnsresearch.ml' as their TLD and first domain, there's a risk that the LLM could identify the tool and recognize that domain as associated with malicious activity. To prevent our model from specifically learning that domain, we remove the TLD '.dnsresearch.ml' from the domain names. We tested our BERT and BERT Hybrid models on the cleaned dataset, taking into account the previously mentioned considerations. Both models achieved 100\% accuracy on the test data, mirroring the same result in training after 10 epochs. This is very optimistic performance.



The primary issue behind the high performance, including 100\% accuracy, observed in some models could be due to data leakage and an overly simplistic dataset. Data leakage occurs when information from outside the training set inadvertently influences model building, such as when training features are directly correlated with the target label. This can lead to artificially inflated performance metrics that do not reflect the model’s true capability 
Additionally, if the dataset is too simple or contains easily identifiable patterns, the model might achieve near-perfect accuracy with minimal challenge, indicating that the dataset does not adequately represent the complexity of the problem domain. 
Based on these findings, it is evident that this dataset may not be suitable for detecting DNS exfiltration attacks, and researchers should consider exploring more complex datasets to better capture the nuances of the problem.

\section{Conclusion}

The development of fine-tuned models for DGA and DNS exfiltration detection provides valuable insights. The fine-tuned model's binary classification results for DGA detection significantly outperform previous work, demonstrating its effectiveness across various datasets, whether small or large, complete or partial. This underscores the importance of fine-tuning in improving detection capabilities, as evidenced by its superior performance in identifying known DGAs compared to earlier methods. However, the model faces challenges in multiclass classification due to the complexity of distinguishing between multiple DGA classes, necessitating further optimization in feature engineering, architecture, and model parameters. Notably, the model excels in detecting unknown DGAs, surpassing state-of-the-art techniques and proving its robustness for real-world cybersecurity scenarios. Despite its success, the analysis of DNS exfiltration datasets reveals that their simplicity leads to near-perfect accuracy with minimal effort, highlighting the need for more complex datasets. In the future, combining multimodal data with LLMs such as DNS, HTTP, and other network traffic may improve the precision of anomaly detection and open the door to more advanced cybersecurity solutions.






%

\bibliographystyle{IEEEtran}
\bibliography{references}

\end{document}